\documentclass{aa}
\usepackage{graphicx}
\usepackage{txfonts}
\usepackage{natbib}
\usepackage{color}
\usepackage[colorlinks, allcolors=blue]{hyperref}

\begin{document}

\title{Oscillations In The Line-of-Sight Magnetic Field Strength In A Pore Observed By The GREGOR Infrared Spectrograph (GRIS)}

\author{C. J. Nelson, R. J. Campbell, M. Mathioudakis}

\offprints{c.nelson@qub.ac.uk}
\institute{Astrophysics Research Centre (ARC), School of Mathematics and Physics, Queen’s University, Belfast, BT7 1NN, N. Ireland, UK.}

\date{}

\abstract
{Numerous magnetohydrodynamic oscillations have been reported within solar pores over the past decades, including in line-of-sight (LOS) velocities, intensities, and magnetic field strengths.}
{Our aim is to identify whether high-amplitude oscillations in the LOS magnetic field strength can be detected within a pore located in Active Region $12748$ and to investigate which physical mechanisms could be responsible for them.}
{A solar pore was observed on the $1$st September $2019$ using the GREGOR Infrared Spectrograph (GRIS) instrument for around one hour. Full-Stokes vectors were sampled in a $37$ \AA\ window containing the \ion{Fe}{I} $15648.52$ \AA\ line (effective Land\'e g-factor of $3$). The LOS magnetic field strength is inferred using the strong-field approximation (SFA). Additionally, the Stokes Inversion based on Response functions (SIR) code is used to gain a more complete understanding the physical properties of the solar atmosphere at the locations of these oscillations.}
{Oscillations of more than $100$ G are observed in the LOS magnetic field in the period window between $600$-$1272$ s at three localised ($>1$\arcsec$^2$) regions. These oscillations have coherence across individual regions indicating that jitter cannot account for their occurrence. Longer-period amplitude variations, amplitudes over $200$ G, are also detected but these have periods outside of the cone-of-influence. Numerical inversions confirm both oscillations in the LOS magnetic field strength at optical depths of around $log\tau_{5000}$=$-0.5$ (potentially caused by compression) and other effects (e.g., changes in the optical depth or the inclination of the magnetic field) may account for these changes.}
{The oscillations in the separations of the Stokes-{\it{V}} lobes of the $15648.52$ \AA\ line appear to be solar in nature. Future work will be required to understand whether these are truly oscillations in the magnetic field strength at a specific depth in the solar atmosphere or whether other effects are responsible for these signatures.}

\authorrunning{C. J. Nelson}
\titlerunning{High-Amplitude Magnetic Field Variations In A Pore}

\maketitle

\section{Introduction}
\label{Introduction}

The detection and analysis of magnetohydrodynamic (MHD) waves within localised magnetic wave-guides in the solar atmosphere has been an incredibly productive topic in solar physics, ever since the initial discovery of such dynamics more than half a century ago (\citealt{Howard68, Beckers69, Wittmann69}). One of the most widely studied,  in the MHD wave context, of the zoo of solar features are pores, which manifest as relatively small (diameters of $1$-$6$ Mm; \citealt{Sobotka03}) and short lived (lifetimes of the order hours to days; \citealt{Sutterlin98}) regions of decreased intensity when compared to the local granulation (\citealt{Bahng58}). Pores make an excellent test-bed for the analysis of MHD waves in the solar atmosphere given that they contain high, kiloGauss (kG), magnetic field strengths (\citealt{Simon70}) and are relatively simple in structure when compared to the larger and more complex sunspots, with their extensive penumbrae (for recent reviews of sunspots see \citealt{Solanki03, Moradi10, Khomenko15}).

\begin{figure*}
\includegraphics[width=0.99\textwidth,trim={0 0 5cm 0}]{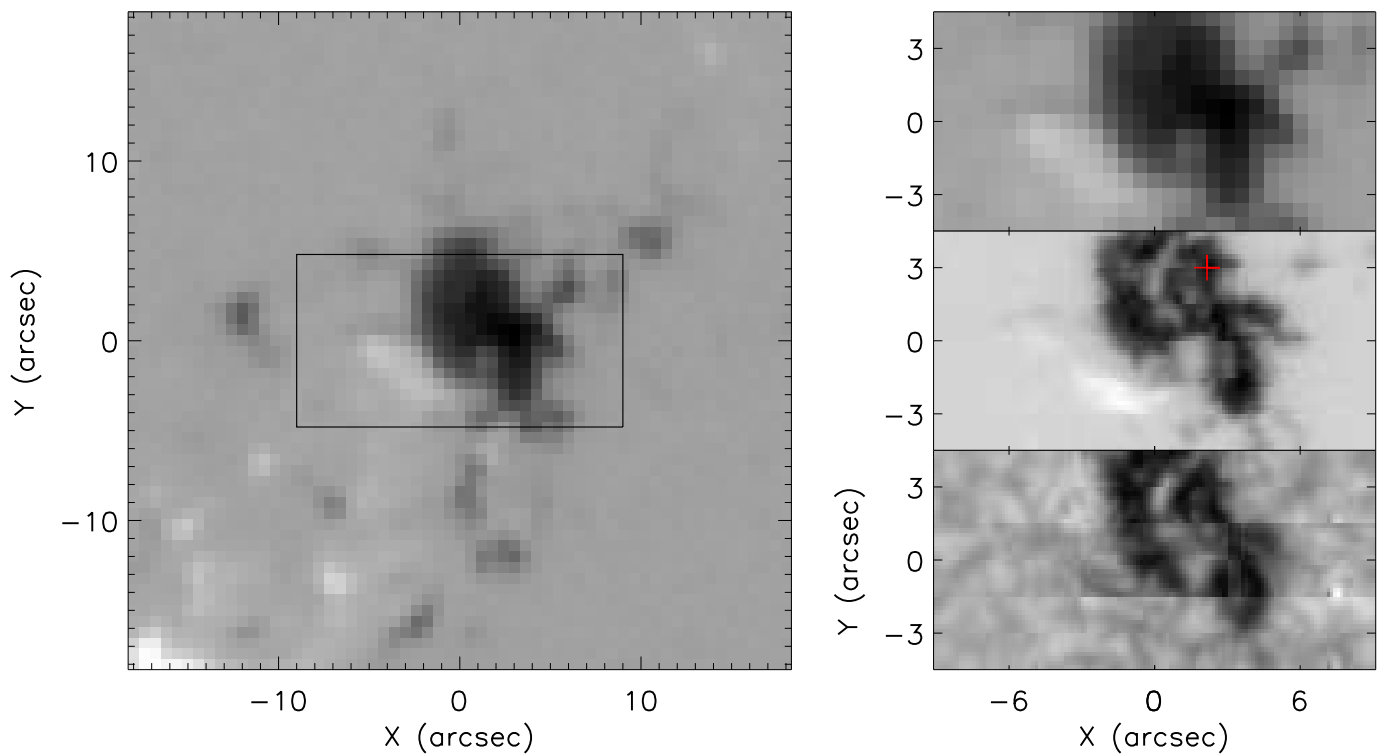}
\caption{(Left panel) Context image of the pore studied in this article as observed by the SDO/HMI instrument at $10$:$00$:$05$ UT on $1$st September $2019$. The over-laid black box outlines the region plotted in the other panels. (Right panels) The total FOV of the GRIS instrument plotted in three different diagnostics: The top row plots a magnetogram returned by the SDO/HMI instrument; the middle row plots the Stokes-{\it{V}} component sampled by GRIS at $15647.6$ \AA; and the bottom row plots the Stokes-{\it{I}} intensity in the continuum at a wavelength position of $15646.8$ \AA. The scan which sampled the GRIS images was started at $10$:$00$:$01$ UT. Artifacts of the stitching together process can be seen in the continuum image. The red cross indicates the pixel discussed in Fig.~\ref{SFA_GRIS}.}
\label{Context_GRIS}
\end{figure*}

Despite being structurally simple, pores are still host to a range of both MHD waves and other dynamic processes (see, for example, \citealt{Hirzberger02, Sobotka13, Freij14, Krishna16, Bharti20, Stangalini21}). Numerous authors have reported on the confident detection of oscillations in a range of observed quantities within pores, including the measured intensities (\citealt{Freij16, Keys18}), areas (\citealt{Morton11, Dorotovic14, Grant15}), and line-of-sight (LOS) velocities (\citealt{Fujimura09, Stangalini12, Cho15}). Additionally, oscillations in the magnetic field strength with root-mean-square amplitudes of $4$-$17$ G have been detected (\citealt{Fujimura09}) highlighting the varied nature of MHD oscillations in pores. Although these quantities are listed separately here, in reality most authors combine measurements of these properties, thereby, allowing a more complete understanding of which MHD mode is responsible for any observed oscillation (\citealt{Moreels13, Moreels15}). 

The majority of oscillations reported in both sunspots and pores have periodicities of around $3$ to $5$ minutes, however, several important examples of long-period oscillations ($>10$ minutes; \citealt{Bakunina13}) have also been identified. \citet{Freij16} observed oscillations in the areas and intensities of two pores with periods of up to $35$ minutes. \citet{Riehokainen19} identified oscillations in pores in the period interval between $18$-$260$ minutes for both the LOS magnetic field strength and intensities sampled by the Solar Dynamics Observatory's Helioseismic and Magnetic Imager (SDO/HMI; \citealt{Scherrer12}) and Atmospheric Imaging Assembly (SDO/AIA; \citealt{Lemen12}) instruments, respectively. More recently, \citet{Stangalini21} identified Alfv\'en (rotational) oscillations of a pore with a fundamental frequency corresponding to around $25$ minutes, as well as several higher-order harmonics. It is known that only two physical mechanisms can be responsible for oscillations in the LOS magnetic field strength at a given height in the solar atmosphere (\citealt{Ulrich96}), namely compression (an increase in the filling factor within the area of one pixel caused by horizontal motions in the solar atmosphere) associated with fast-mode waves and bending (where the orientation of the field lines with respect to the observer changes through time) associated with Alfv\'en waves. Additionally, changes in the observed optical depth could also cause oscillations in the measured LOS magnetic field strength due to gradients of the magnetic field (around $4$ G km$^{-1}$ in pores; \citealt{Sutterlin98}). The presence of long-period oscillations in both smaller magnetic structures such as network elements (see, for example, \citealt{Kolotkov17}) and sunspots  (see Table 1 of \citealt{Grinon20} for a comprehensive overview) are also well established with detections having been made in a large range of diagnostics.

Our understanding of the vast array of oscillations within both pores and sunspots is further complicated by the presence, from the photosphere to the corona, of both spatially local (diameters of below $1$\arcsec\ in sunspots; see, for example, \citealt{Centeno05, Henriques20}) and global wave signatures. A range of explanations such as $p$-mode penetration (\citealt{Felipe17}), chromospheric resonators (\citealt{Hollweg79, Jess20}), and slow-wave subphotospheric resonators (\citealt{Zhugzhda14, Zhugzhda18}) have been proposed to account for these rich spectrum of oscillations, however, no single model is seemingly able to account for the gamut of observed wave signatures making this a topic of active and vigorous debate in the community. A more complete understanding of long-period ($>10$ minutes) oscillations, both local and global, within pores and sunspots would potentially help to differentiate between these models in the future (\citealt{Zhugzhda18}).

In this article, we investigate spectro-polarimetric data sampled by the Intergral Field Unit (IFU) of the GREGOR Infrared Spectrograph (GRIS; \citealt{Collados12}) in order to gain in-sight into long-period, high-amplitude oscillations of the LOS magnetic field strength in a solar pore. Our work is structured as follows: In Sect.~\ref{Methods} we detail the observations analysed in this work and the techniques used to analyse them; In Sect.~\ref{Results} we present the results of our analysis; In Sect.~\ref{Discussion} we provide a discussion of our results in the context of current literature; before, finally, in Sect.~\ref{Conclusions} we draw our conclusions.

\begin{figure*}
\includegraphics[width=0.99\textwidth]{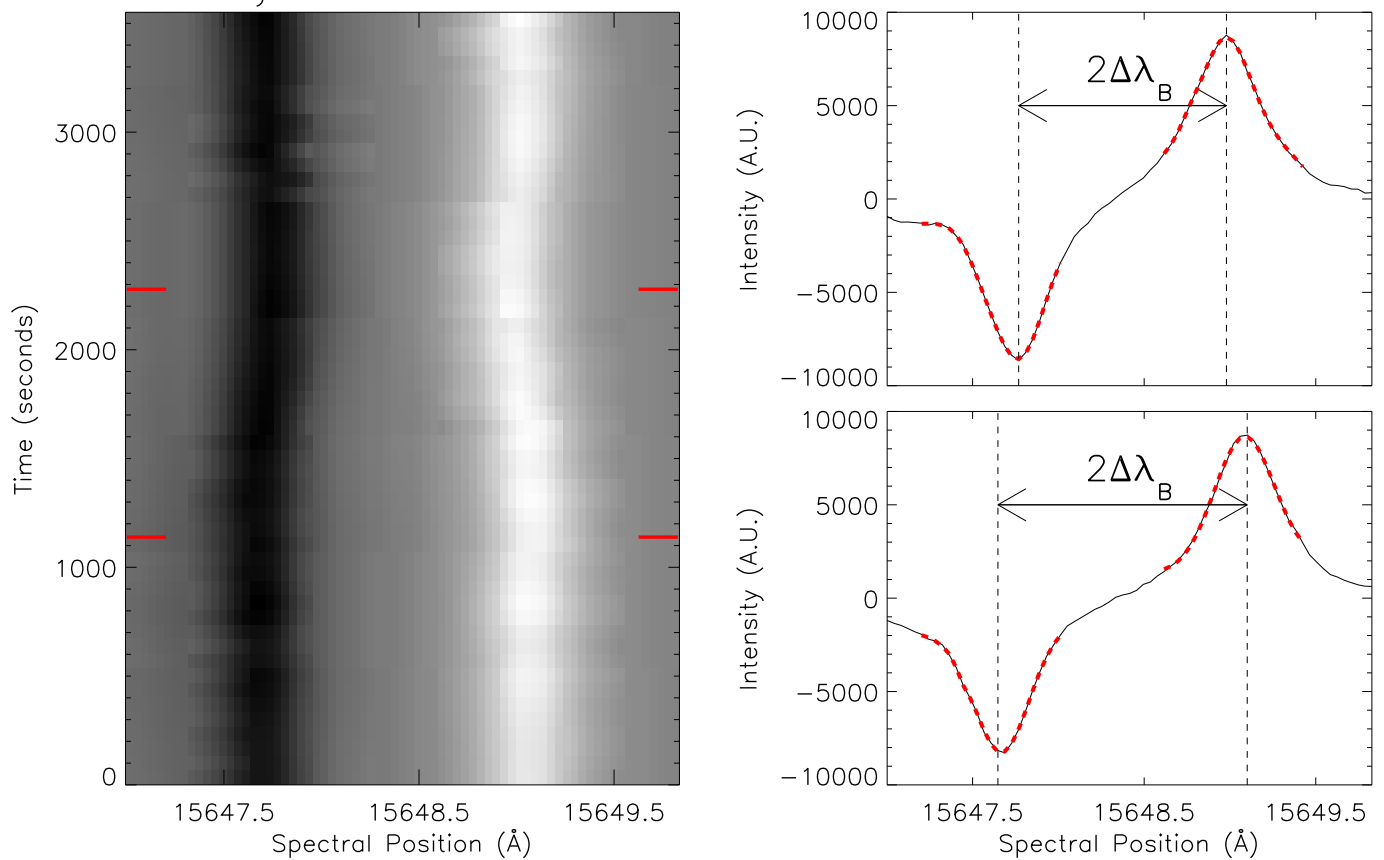}
\caption{(Left panel) Stokes-{\it{V}} spectral-time plot for a window around the \ion{Fe}{I} $15648.52$ \AA\ line sampled at the pixel denoted by the red cross on Fig.~\ref{Context_GRIS}. The SFA allows us to infer the LOS magnetic field strength purely through measuring the distance between peaks of the positive and negative lobes. (Right panels) Stokes-{\it{V}} plots measured at the locations marked by the upper (top panel) and lower (bottom panel) red tick marks over-laid on the left panel. The vertical dashed lines indicate the minima or maxima of the lobes. The dashed red curves plot the Gaussian fits, within a $0.8$ \AA\ window used to calculate the locations of the minima or maxima. The lobe widths in the top and bottom panels correspond to $B_\mathrm{SFA}$ values of $1750$ G and $2100$ G, respectively.}
\label{SFA_GRIS}
\end{figure*}

\section{Methods}
\label{Methods}

\subsection{Observations}

The ground-based data analysed in this article were acquired with the GRIS instrument at the $1.5$ m GREGOR solar telescope (\citealt{Schmidt12, Kleint20}) between $10$:$00$:$01$ UT and $10$:$58$:$38$ UT on the $1$st September $2019$. After this time the telescope was pointed to a different target. During these observations, GRIS conducted a $3\times3$ mosaic over the lead pore of AR $12748$ in double sampling mode. The pore was located at $\mu=\cos\theta=0.87$, where $\theta$ is the heliocentric angle. A spectral window spanning between $15639.291$ \AA\ and $15676.553$ \AA\ was measured for each of the four polarimetric states (I, Q, U, and V) with a spectral sampling of approximately $0.04$ \AA, corresponding to around $0.8$ km s$^{-1}$ in velocity units. The use of double sampling in the spatial directions allowed a field-of-view (FOV) of around $18$\arcsec$\times9$\arcsec\ to be observed with a cadence of close to $67$ s (four $30$ ms exposures per polarimetric state at each pointing position) and pixel scales of $0.135$\arcsec\ and $0.1875$\arcsec\ in the $x$- and $y$-directions, respectively. Data reductions were completed using the standard GRIS pipeline, which included dark and flat field corrections and the removal of cross-talk between polarimetric states. In addition to the ground-based data, LOS magnetograms sampled by SDO/HMI were also studied to provide context about the pore and the wider AR during this time-period. These SDO/HMI data have a post-reduction pixel scale of $0.6$\arcsec\ and a cadence of $45$ s.

Several interesting lines are present within the GRIS spectral window studied here including the photospheric \ion{Fe}{I} $15648.52$ \AA\ and $15652.87$ \AA\ lines with effective Land\'e $g$-factors, $g_\mathrm{eff}$, of $3$ and $1.5$ (\citealt{Borrero03}), respectively. The high magnetic sensitivity of these lines means these data are excellent for studying the evolution of the photospheric magnetic field within the pore over time-scales of the order minutes. In Fig.~\ref{Context_GRIS} we plot an overview of the FOV analysed in this article for reference.  The left-hand panel plots a context $36$\arcsec$\times36$\arcsec\ image of the pore studied here as observed by the SDO/HMI instrument at $10$:$00$:$05$ UT on $1$st September $2019$. This pore was relatively small, allowing it to be almost completely observed by the GRIS FOV denoted by the black box. The right-hand panels plot the FOV of GRIS co-temporally in three different diagnostics, namely: A zoom-in of the SDO/HMI magnetogram plotted in the left-hand panel (top); the Stokes-{\it{V}} intensity sampled by GRIS at $15647.6$ \AA\ (middle); and the Stokes-{\it{I}} intensity sampled by GRIS at a continuum position of $15646.8$ \AA\ (bottom). Artifacts of the stitching together of multiple panels to create the total GRIS FOV are evident in the bottom panel, however, we confirmed that the signatures studied here were not confined to only such locations. The red cross over-laid on the middle panel indicates the pixel studied in Fig.~\ref{SFA_GRIS}.

\subsection{Strong-Field Approximation}

\begin{figure*}
\includegraphics[trim={6cm 0 0 0},width=0.99\textwidth]{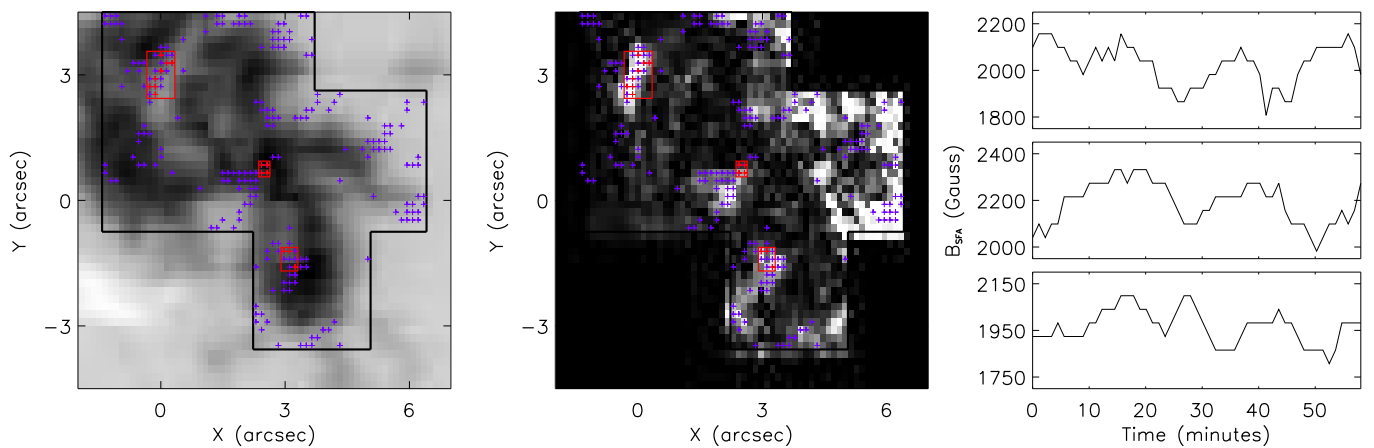}
\caption{(Left panel) Stokes-{\it{V}} map at $15647.6$ \AA\ at the same time as plotted in Fig.~\ref{Context_GRIS}. The black contour outlines the pixels which were analysed using the wavelet method. The red crosses indicate the $15$ pixels which displayed sustained high-amplitude oscillations within the period range studied here. Purple crosses locate all other pixels which return some periodicity, albeit with either low amplitude or localised in time. The red boxes indicate the spatial regions with sustained oscillations which are studied in Sect.~\ref{Wavelet}. (Centre panel) Total wavelet power within the cone-of-influence around a period of $\sim931$ s for each pixel within the black contour plotted on the left-hand panel. White indicates more power, with black indicating relatively less power. The three regions of sustained power studied here are immediately evident in the centre of the pore. (Right panels) $B_\mathrm{SFA}$ against time for three of the pixels indicated by the red crosses in the left-hand panel. These three pixels (Pixels $2$, $9$, and $15$ from top to bottom; see Table~\ref{Stats_GRIS}) were selected such that one came from each of the three red boxes, with the position of the plots corresponding to the position of the boxes (top, middle, and bottom, respectively). Long-period, high-amplitude oscillations in the inferred LOS magnetic field strength are evident in each panel.}
\label{Wavelet_GRIS}
\end{figure*}

In this article, we use the Strong-Field Approximation (SFA) to infer the presence of any oscillations within the observed pore. This technique, essentially, allows us to infer the strength of the LOS magnetic field, $B_\mathrm{SFA}$, through the measurement of the Zeeman splitting between the two lobes in the Stokes-{\it{V}} component of the observed profiles. Using the SFA, $B_\mathrm{SFA}$ can be calculated using the formula (see, for example, \citealt{Ulrich96}):
\begin{equation}
B_\mathrm{SFA}(G)\approx\frac{10^{13}\times\Delta\lambda_\mathrm{B}(\AA)}{4.67\times{g}_\mathrm{eff}\times\lambda^2_\mathrm{0}(\AA)}
\label{SFA_eq}
\end{equation}
where $\Delta\lambda_\mathrm{B}$ is half the separation between the peaks of the lobes measured from the observed Stokes-{\it{V}} profiles and $\lambda_\mathrm{0}$ is the central wavelength of the studied line. Limiting our analysis to the most appropriate spectral line in the observed wavelength window returns $\lambda_\mathrm{0}$=$15648.52$ \AA\ and $g_\mathrm{eff}$=$3$. Given the spectral sampling of these data ($0.04$ \AA), a change of one pixel in the separation between the two Stokes-{\it{V}} lobes would lead to a change of around $58$ G in the returned value of $B_\mathrm{SFA}$.

In Fig.~\ref{SFA_GRIS}, we plot an example of the application of the SFA, both for completeness and to demonstrate the suitability of this method for these data. In the left-hand panel, we plot a Stokes-{\it{V}} spectral-time plot for the pixel indicated by the red cross over-laid on Fig.~\ref{Context_GRIS}. Clear variations in the separations of Stokes-{\it{V}} lobes are evident with time indicating that potential variations in $B_\mathrm{SFA}$ are present. The right-hand panels of Fig.~\ref{SFA_GRIS} more clearly display these variations with time, with the top panel plotting the Stokes-{\it{V}} profiles at their approximate weakest point and the bottom panel plotting the Stokes-{\it{V}} profiles at their approximate strongest point. The top and bottom panels correspond to the upper and lower tick marks on the left-hand panel, respectively. Both the positive and negative lobes of the Stokes-{\it{V}} profiles are fitted with Gaussians (red dashed curves) in a $0.8$ \AA\ spectral window, with the separation between the peaks of these Gaussians (denoted by the black dashed vertical lines) indicating the spectral width, $2\Delta\lambda_\mathrm{B}$. Inserting these measured values of $\Delta\lambda_\mathrm{B}$ into Eq.~\ref{SFA_eq} gives LOS magnetic field strengths of around $1750$ G for the top panel and $2100$ G for the bottom panel. The SFA was applied to each pixel within the region denoted by the black contours on the left-hand panel of Fig.~\ref{Wavelet_GRIS} at each time-step within this dataset.

\subsection{Inversions}

\begin{table}[b]
\caption{Number of free parameters (nodes) used in relevant atmospheric parameters in S1 and S2.}       
\label{table:nodes}      
\centering                          
\begin{tabular}{| c | c |c | c |c | c |c | c | c |}    
%\cline{2-5}
\hline 
  Parameter &$T$ & $v_{mic}$ & $v_{LOS}$ & $v_{mac}$ & $B_\mathrm{SIR}$ & $\gamma$ & $\Phi$ & $\alpha$\\
 \hline                 

  Nodes in S1 & 3 & 1 & 2 & 1 & 2 & 2 & 1 & 0\\
\hline
  Nodes in S2 & 3 & 1 & 2 & 1 & 0 & 2 & 1 & 0\\
\hline
\end{tabular}
\end{table}

In order to investigate the thermodynamic, kinematic and magnetic properties of the atmosphere at the locations where potential oscillations were identified, we also analyse these data using the least-squares Stokes Inversions based on Response functions (SIR) code \citep{SIR}. We adopt a similar approach to \cite{Campbell2021} in that we repeat the inversion procedure a number of times per pixel with randomized model parameters, namely: the inclination of the magnetic field with respect to the LOS, $\gamma$; the azimuth of the magnetic field in the plane-of-sky, $\Phi$; the LOS magnetic field strength, $B_\mathrm{SIR}$; and the LOS velocity, $v_{\mathrm{LOS}}$. This reduces the probability of converging to a solution that is located in a local, as opposed to global, minimum in the $\chi^2$ hyper-surface. For the remaining atmospheric parameters (namely: temperature, T; electron pressure; gas density, $\rho$; and gas pressure) we initialise the inversions with a hot umbral model, which prescribes the stratification of the atmosphere for each of these parameters against the log of the optical depth at $5000$ \AA, log$\tau_{5000}$. We assume the magnetic filling factor, $\alpha$, is $1$. Additionally, we include a static $30\%$ global unpolarised stray light component as measured in the surrounding quiet Sun. 

We allow the macroturbulent velocity, $v_{\mathrm{mac}}$, which accounts for the presence of unresolved, turbulent motions within the resolution element, to vary as a free parameter. We do not include an estimated spectral point spread function (PSF), but instead allow $v_{\mathrm{mac}}$ to encapsulate this effect. The microturbulent velocity, $v_{\mathrm{mic}}$, which accounts for small-scale turbulent velocities, is also included as a free parameter but is assumed to be constant in depth. Table \ref{table:nodes} summarises the nodes used in both the first (S1) and second (S2) inversion schemes for all relevant parameters. The notable differences between S1 and S2 are the number of repeats ($150$ for S1 and $50$ for S2) and that $B$ is fixed (based on the values from the first frame) in S2 to allow us to investigate whether variations in other atmospheric parameters can account for the observed changes in the Stokes vector through time. Essentially, S1 will allow us to investigate whether compression could account for these signatures whilst S2 will allow us to test whether bending or changes in the optical depth could be responsible. We use the same atomic data as used by \cite{Campbell2021} and invert the same five spectral lines. By inverting multiple lines, with varying strengths, opacities, and $g_{\mathrm{eff}}$ values, we can probe the atmosphere at a greater range of heights than if we inverted only one line. \citet{Campbell21b} used MHD simulation outputs to evaluate the extent to which the atmospheric parameters may be recovered by SIR  inversions in the context of the quiet Sun. The authors found that high correlations between MHD and inverted values are found deep in the atmosphere for $T$, $v_{\mathrm{LOS}}$, $\gamma$ and $B$, with the kinetic and thermodynamic parameters sensitive to deeper layers than the magnetic parameters.

\section{Temporal Magnetic Field Variations}
\label{Results}

\subsection{Wavelet Analysis}
\label{Wavelet}

\begin{table}
\begin{center}
\begin{tabular}{|c|c|c|c|c|c|}
 \hline
Pixel & x & y & Region & Period ($>$$99$ \%) & Amplitude \\
 \hline
1 & 65 & 6 & Upper & $>$$800$ s & $110$ G \\
2 & 66 & 6 & Upper & $>$$800$ s & $130$ G \\
3 & 67 & 5 & Upper & $>$$800$ s & $210$ G \\
4 & 67 & 7 & Upper & $800$-$1100$ s & $160$ G \\
5 & 68 & 8 & Upper & $>$$800$ s & $200$ G \\
6 & 68 & 9 & Upper & $>$$800$ s & $180$ G \\
7 &  68 & 10 & Upper & $800$-$1200$ s & $180$ G \\
8 & 69 & 9 & Upper & $>$$800$ s & $200$ G \\
\hline
9 & 48 & 19 & Central & $>$$1100$ s & $140$ G \\
10 & 48 & 20 & Central & $>$$1050$ s & $160$ G \\
11 & 49 & 19 & Central & $>$$1100$ s & $110$ G \\
12 & 49 & 20 & Central & $>$$1000$ s & $170$ G \\
\hline
13 & 43 & 32 & Lower & $>$$800$ s & $120$ G \\
14 & 44 & 30 & Lower & $800$-$1000$ s & $110$ G \\
15 & 45 & 30 & Lower & $600$-$1000$ s & $120$ G \\
 \hline
\end{tabular}
\end{center}
\caption{Properties of the 15 pixels indicated by the red crosses on Fig.~\ref{Wavelet_GRIS}. The $x$ and $y$ columns correspond to the original coordinates of the data (pre-alignment with SDO/HMI), the `Region' column details which red box from Fig.~\ref{Wavelet_GRIS} the pixel is within, the `Period' column denotes the approximate temporal ranges with significance above the $99$ \%\ confidence interval (max period is $1272$ s), and the `Amplitude' column is an estimate of the maximum peak-to-peak value. Errors for the periods and peak amplitudes are difficult to quantify (due to, for example, the varying amplitudes of the apparent oscillations through time, the snap-shot nature of these observations, and noise in the data), however, we are confident that these reported values are within $\pm50$ s and $\pm20$ G of their true values, respectively.}
\label{Stats_GRIS}
\end{table}

\begin{figure}
\includegraphics[width=0.99\columnwidth]{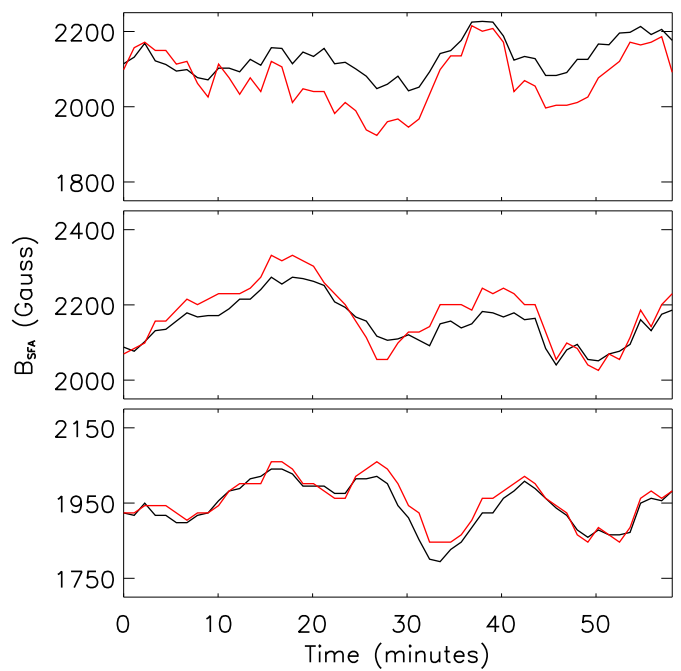}
\caption{Evolution of the LOS magnetic field strength averaged over multiple pixels for each of the upper (top panel), central (middle panel), and lower (bottom panel) red boxes over-laid on Fig.~\ref{Wavelet_GRIS}. The red curves plot the magnetic field averaged over the red crosses within each box and the black curves plot the average magnetic field across the entire red box ($\pm$ one pixel either side for the central box). These apparent oscillations are clearly sustained and in-phase across multiple pixels in these data.}
\label{Average_GRIS}
\end{figure}

We begin our analysis by identifying oscillations in $B_\mathrm{SFA}$ that are confidently detected using signal analysis methods. The Morlet wavelet (based on \citealt{Torrence98}) from the in-built IDL wavelet package is used and a $99$ \% confidence level is selected. We study the temporal evolution of the magnetic field strength within $1739$ pixels located inside the region in the pore outlined by the black contour on the left-hand panel of Fig.~\ref{Wavelet_GRIS}. Only pixels where periods between $600$ s (the lower limit for long-period oscillations defined by \citealt{Bakunina13}) and $1272$ s (the maximum value allowed by the cone-of-influence) are present were selected as potential candidates for study here. Overall, $233$ pixels ($13.3$ \%) display signals above the $99$ \% confidence level in the studied temporal range, with each of these pixels being denoted by a coloured cross on Fig.~\ref{Wavelet_GRIS}. As we are interested in the detection of long-period and high-amplitude oscillations, we limit our analysis in this article to pixels which satisfy three criteria: 1) The pixel is sufficiently far from the edge of the pore that any evolution in area will not produce artificial magnetic field changes (induced by the movement of the edge of the pore); 2) the amplitude of the oscillation is higher than the average standard deviation of $B_\mathrm{SFA}$, estimated to be approximately $90$ G (calculated from all pixels studied); and 3) the oscillation is sustained for multiple periods. The application of these criteria returned $15$ pixels (red crosses), located within three sub-regions denoted by the red boxes on Fig.~\ref{Wavelet_GRIS}, which will be studied in more detail here. All other pixels which have potential oscillatory signatures are denoted by the purple crosses on Fig.~\ref{Wavelet_GRIS}. The centre panel of Fig.~\ref{Wavelet_GRIS} plots the summed power within the cone-of-influence returned by the Wavelet analysis around a period of $\sim931$ s, with the regions containing the sustained oscillations studied here being immediately evident within the centre of the pore. The other bright regions to the right-hand side of the pore appear to be outside of the magnetic structure itself and, as such, were not studied here. Estimates for the periods, with $99$ \%\ significance, and amplitudes of the oscillations of $B_\mathrm{SFA}$ at each of the $15$ pixels studied in this article are reported in Table~\ref{Stats_GRIS}.

\begin{figure*}
\includegraphics[width=0.99\textwidth]{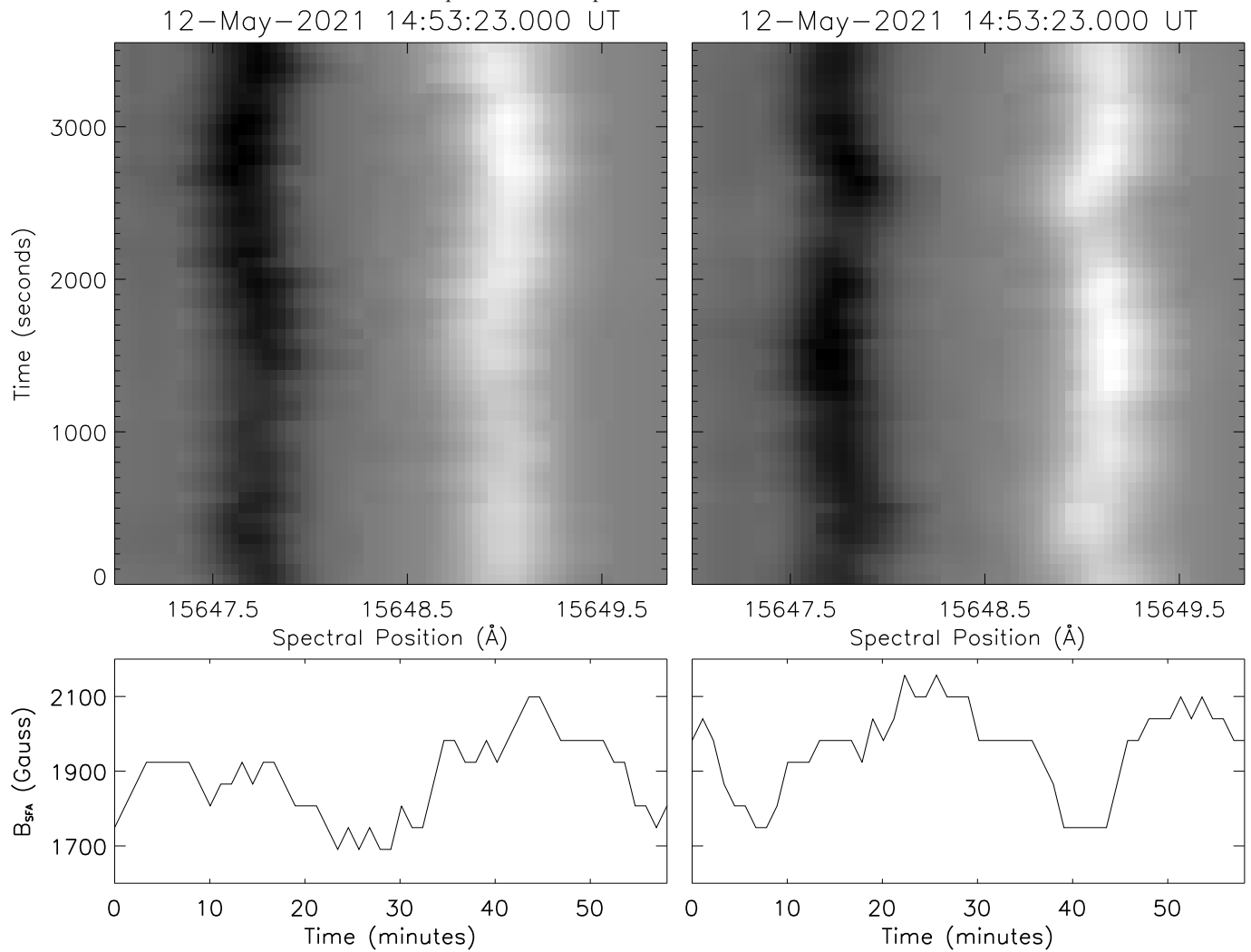}
\caption{Two examples of longer-period amplitude variations present in these data. (Top panels) Spectral-time plots constructed using Stokes-{\it{V}} data in the same spectral window as plotted in the left-hand panel of Fig.~\ref{SFA_GRIS}. (Bottom panels) Variation of $B_\mathrm{SFA}$ through time calculated from the Stokes-{\it{V}} data plotted in the top panels. Peak-to-peak amplitude variations are measured to be more than $400$ G in both of these cases.}
\label{Variations_GRIS}
\end{figure*}

In the right-hand panels of Fig.~\ref{Wavelet_GRIS}, we plot the evolution of $B_\mathrm{SFA}$ against time for three of the pixels indicated by the red crosses in the left-hand panel. The pixels plotted in the top (Pixel $2$), middle (Pixel $9$), and bottom (Pixel $15$) panels are representative of the oscillations present in the upper, central, and lower red boxes over-laid on the left-hand panel, respectively. Wavelet analysis applied to Pixel $2$ indicates significance at the $>99$ \%\ level at periodicities $>800$ s. For Pixel $9$, $>99$ \%\ significance is found for periods $>1100$ s. Finally, wavelet analysis applied to Pixel $15$ returns $>99$ \%\ significance in the period window between $600$-$1000$ s. The amplitudes of the apparent oscillations of $B_\mathrm{SFA}$ in each of these pixels is in the range of $120$-$140$ G during this time-period. Higher amplitudes are also present within these $15$ locations, with a maximum value of $210$ G being recorded. We note that the amplitudes of these oscillations are not constant through time (as can clearly be seen in Fig.~\ref{Wavelet_GRIS}), with the estimates reported in Table~\ref{Stats_GRIS} being calculated using the maximum peak-to-peak variations in $B_\mathrm{SFA}$.

Each of the three regions in the centre of the pore which contain the most confidently detected periodic signatures are relatively small (being less than $1$\arcsec$^2$) meaning it is important to investigate whether any jitter, slight mis-alignments between frames, which could move regions of high LOS magnetic field strength in and out of individual pixels, remaining in the dataset after reduction could have introduced these apparent oscillations. In order to investigate whether jitter was present, we applied cross-correlation techniques to consecutive frames in both Stokes-{\it{I}} continuum images and a small ([60,27] pixel) region of the LOS magnetic field strength inferred using the SFA. No large or systematic shifts were evident indicating that jitter is not responsible for these apparent oscillations. The spatial coherence of these oscillations is displayed in Fig.~\ref{Average_GRIS}, where the temporal evolution of two distinct measures for each of the upper (top panel), central (middle panel), and lower (bottom panel) red boxes over-laid on Fig.~\ref{Wavelet_GRIS} are plotted. The red lines plot the $B_\mathrm{SFA}$ values through time averaged over all of the red crosses in the respective box whilst the black solid line plots the mean $B_\mathrm{SFA}$ value constructed by averaging the pixels within the entire red box ($\pm$ one pixel each side for the central region). Clearly the oscillations appear to be in-phase across a number of pixels providing further confidence that sub-pixel jitter is not responsible for these apparent oscillations.

\begin{figure*}
    \centering
    \includegraphics[width=\textwidth]{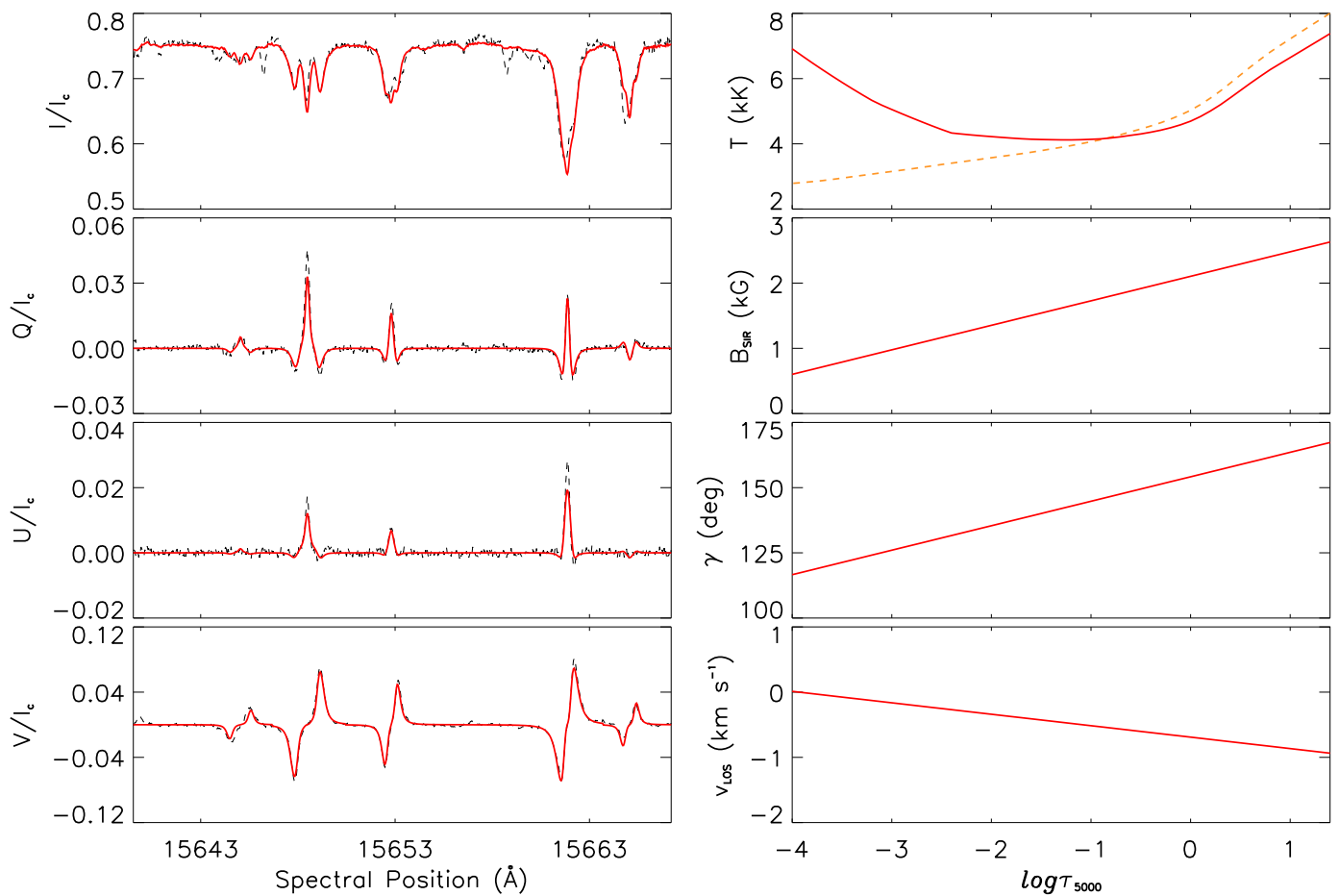}
    \caption{(Left column) The full continuum-normalised Stokes vector ($I$, $Q$, $U$, and $V$ from top to bottom) is shown for all of the five spectral lines considered by the inversion scheme for Pixel $15$ from Table \ref{Stats_GRIS} ($x$=$45$, $y$=$30$). The dashed black lines plot the observed spectra and the solid red lines plot the fitted spectra returned by SIR with S1. (Right column) The retrieved atmospheric parameters (namely $T$, $B_\mathrm{SIR}$, $\gamma$ and $v_{\mathrm{LOS}}$ from top to bottom) are shown as a function of log$\tau_{5000}$. The stratification of $T$ in the initial hot umbral model is also shown (orange, dashed line over-laid on the top right panel). The value of $\Phi$ during this inversion cycle was $96^\circ$.}
    \label{fig:inversion_example}
\end{figure*}

To complete our analyis of the apparent oscillations of $B_\mathrm{SFA}$, we next investigated whether seeing effects could be responsible for their occurrence in this dataset. The periodicity, amplitude, and spatial structuring within the pore gives us confidence that changes in the Fried's parameter, $r_\mathrm{0}$, with time are not responsible for these apparent oscillations, however, we did investigate this through analysis of the $r_\mathrm{0}$ value recorded by the telescope at each time-step (not plotted here). No long-period oscillation was detected, as predicted, indicating that these apparent oscillations are not caused by changes in the seeing conditions at the telescope through time. We also correlated $B_\mathrm{SFA}$ and the root mean square continuum intensity contrast in each frame, defined as
\begin{equation}
    \delta I_{\mathrm{rms}} \approx \frac{\sqrt{\frac{1}{N}\sum_{x,y}\bigg(I(x,y) - \langle I_{x,y}\rangle \bigg)^2}}{\langle I_{x,y}\rangle},
\end{equation}
where $I(x,y)$ are continuum intensity values of individual pixels and $\langle I_{x,y}\rangle$ is the mean over all pixels, $N$. This parameter better accounts for all the residual aberrations at the focal plane than the $r_0$. No significant correlations were detected between this quantity and any of the $15$ pixels studied. Additionally, these oscillations are not limited to one single GRIS tile nor limited to the locations where multiple tiles are stitched together indicating that they are not caused by issues in the reduction pipeline. Overall, we can conclude that these apparent oscillations do not appear to be introduced by instrumental or seeing effects and are, therefore, likely solar in nature. Further discussion about whether they are real variations of the LOS magnetic field strength or are a result of changes in the other atmospheric parameters will be investigated in more detail using inversion techniques in Sect.~\ref{SIR}.

\subsection{Longer-Period Amplitude Variations}
\label{Variations}

In addition to the oscillations confidently detected in the $600$-$1272$ s window through wavelet analysis, several pixels display large amplitude variations in $B_\mathrm{SFA}$ with apparent periods longer than the maxima of the cone-of-influence. As such, we do not call these signatures oscillations but instead refer to them as longer-period amplitude variations. One example of a pixel where longer-period amplitude variations are present in $B_\mathrm{SFA}$ is plotted in Fig.~\ref{SFA_GRIS}, where the apparent period is around $2500$ s, approximately double what can be confidently detected through signal analysis methods with these data. In the top two panels of Fig.~\ref{Variations_GRIS}, we plot spectral-time plots for two further pixel in the GRIS FOV where longer-period amplitude variations in $B_\mathrm{SFA}$ are present. Temporal changes in the widths in the separation between the two Stokes-{\it{V}} lobes are immediately evident with apparent periodicities between $1500$-$2000$ s. We note that the apparent periods reported for the three longer-period amplitude variations plotted here should not be over-interpreted, with each event being selected purely to demonstrate the general behaviour observed within the pore. It is certainly possible, for example, that variations with even longer apparent periods may be present if a dataset sampled over an extended time-period of several hours were to be studied in the future.

\begin{figure*}
    \centering
    \includegraphics[width=.99\textwidth]{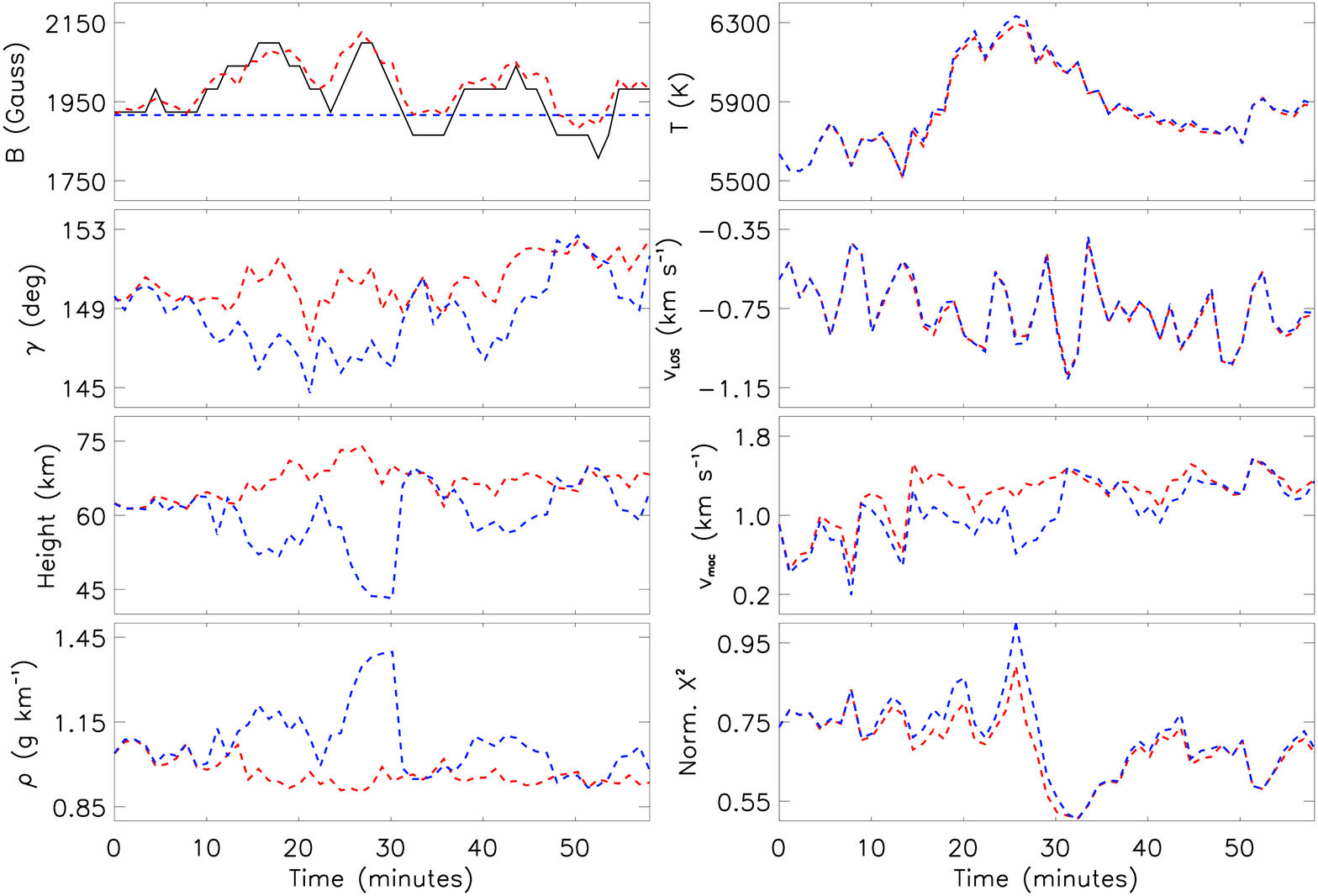}
    \caption{The evolution of various atmospheric parameters ($B_\mathrm{SIR}$; $T$; $\gamma$; $v_\mathrm{LOS}$; the optical depth; $v_\mathrm{mac}$; and $\rho$) returned by the S1 (red lines) and S2 (blue lines) inversion schemes for the full time-series at appropriate optical depths for Pixel $15$. The bottom right panel plots the normalised $\chi^2$ fits for both inversion schemes, displaying neither appears to fit these spectra significantly better than the other. The evolution of $B_\mathrm{SFA}$ is also over-laid on the top left panel (black line). Note that $B_\mathrm{SFA}$ and the LOS magnetic field strength returned by S1 match well, whereas the LOS magnetic field strength does not change through time in the S2 inversions. All inverted parameters are plotted at $log\tau_{5000}$=$-0.5$, except for $T$ which is plotted at $log\tau_{5000}$=$0.5$.}
    \label{fig:temporal_plots}
\end{figure*}

Similar to the shorter-period oscillations discussed in the Sect.~\ref{Wavelet}, these longer-period amplitude variations can also display large deviations in $B_\mathrm{SFA}$. The pixel plotted in Fig.~\ref{SFA_GRIS}, for example, displays an apparent amplitude of around $175$ G. In the bottom two panels of Fig.~\ref{Variations_GRIS}, we plot the evolution of $B_\mathrm{SFA}$ through time for the pixels plotted in the upper panels. These amplitude variations both have peak-to-peak changes of over $400$ G, giving an apparent amplitude of more than $200$ G at these locations. These values are close to the maximum values reported for the $15$ pixels studied in the previous section (reported in Table~\ref{Stats_GRIS}). Given seeing conditions often limit the acquisition of data using ground-based instruments to time-scales of around one hour, it is likely that space-borne data will be required in the future to better understand amplitude variations with time-scales such as those presented here, and potentially even longer.

\subsection{Inversions}
\label{SIR}

To better understand which physical mechanisms could be responsible for the observed apparent oscillations in the LOS magnetic field strength, we also invert the Stokes profiles observed at these locations using the SIR code. We note that these inversions will not allow us to answer the question of which specific mechanism is responsible for these oscillations but it will, instead, allow us to test whether specific hypotheses are consistent and worthy of further study. In the left-hand column of Fig.~\ref{fig:inversion_example}, we plot sample Stokes profiles ($I$, $Q$, $U$, and $V$ from top to bottom) from Pixel 15 from Table~\ref{Stats_GRIS} ($x$=$45$, $y$=$30$) with black dashed lines, and the corresponding synthetic profiles produced by SIR using S1 with solid red lines. The inversion is able to fit the observed Stokes profiles qualitatively well. Specifically, the large Zeeman splitting in both Stokes-{\it{I}} and {\it{V}} is well reproduced. The red lines in the right-hand panels of Fig.~\ref{fig:inversion_example} plot the atmospheric parameters ($T$, $B_
\mathrm{SIR}$, $\gamma$, and $v_\mathrm{LOS}$) returned by SIR to reproduce the observed spectra against optical depth. The dashed orange line in the temperature panel plots the hot umbral model used to initiate the inversions. The value of $B_\mathrm{SFA}$ for this pixel at this time was $1923$ G whilst $B_\mathrm{SIR}$, returned from the S1 inversion, at $log\tau_{5000}$=$-0.5$ (the peak of the response function for the $15648.52$ \AA\ line) was $1916$ G, showing very close agreement. It is important to note that as these observations were taken at $\mu = 0.87$, and SIR returns magnetic and velocity vectors with respect to the LOS, a $\gamma$ value approximately of $150^{\circ}$ is vertical in the solar atmosphere (as opposed to $0^{\circ}$ or $180^{\circ}$). 

In Fig. \ref{fig:temporal_plots}, we plot the evolution of each of the atmospheric parameters returned by both the S1 (red lines) and S2 (blue lines) inversion schemes, at specific optical depths, through time for Pixel 15 from Table~\ref{Stats_GRIS}. For S1, the evolution of $B_\mathrm{SIR}$ (top left panel) at an optical depth of $log\tau_{5000}$=$-0.5$ matches well with $B_\mathrm{SFA}$ (black line). The value of $T$ (top right panel) at $log\tau_{5000}$=$0.5$ from S1 evolves from around $5600$ K at the beginning of the time-series to reach a peak above $6200$ K at $25$ min, before dropping to below $6000$ K at the end of the time-series. Importantly, though, no oscillation is detected in the temperature during this time. Additionally, no oscillation with strong correlation to $B_\mathrm{SFA}$ is present within the $\gamma$ value returned by S1 (second panel from top on the left) which remains relatively stable at around $150^\circ$ during this time-period. The $5$ minute $p-$mode oscillation (see, for example, \citealt{Lites82, Khomenko15}) in $v_{\mathrm{LOS}}$ is clearly evident (second panel down on the right side) from S1 with an amplitude of up to $0.4$ km s$^{-1}$, however, no longer-period ($>600$ s) oscillation is present. The reproduction of the clear $5$ minute periodicities gives us confidence that the inversions are accurately capturing the dynamics of the solar atmosphere. The third row of Fig.~\ref{fig:temporal_plots} plots the changes of the height scale (left panel), caused predominantly by the variations in the temperature, and $v_\mathrm{mac}$ (right panel). In SIR, the height scale is defined so that $0$ km is at $log\tau_{5000}$=$0.0$, meaning the height at $log\tau_{5000}$=$-0.5$ is defined relative to this layer. At the time of the minimum temperature, we observe approximately $10$ km deeper in the photosphere than when temperature is at its greatest. Notably, however, neither the height scale nor $v_\mathrm{mac}$ display any long-period ($>600$ s) oscillation. The bottom row of Fig.~\ref{fig:temporal_plots}, plots the variation of $\rho$ and the $\chi^2$ fits from SIR. Again, neither displays any long-period oscillation which could account for the oscillation in $B_\mathrm{SIR}$. 

\begin{figure}
    \centering
    \includegraphics[width=\columnwidth]{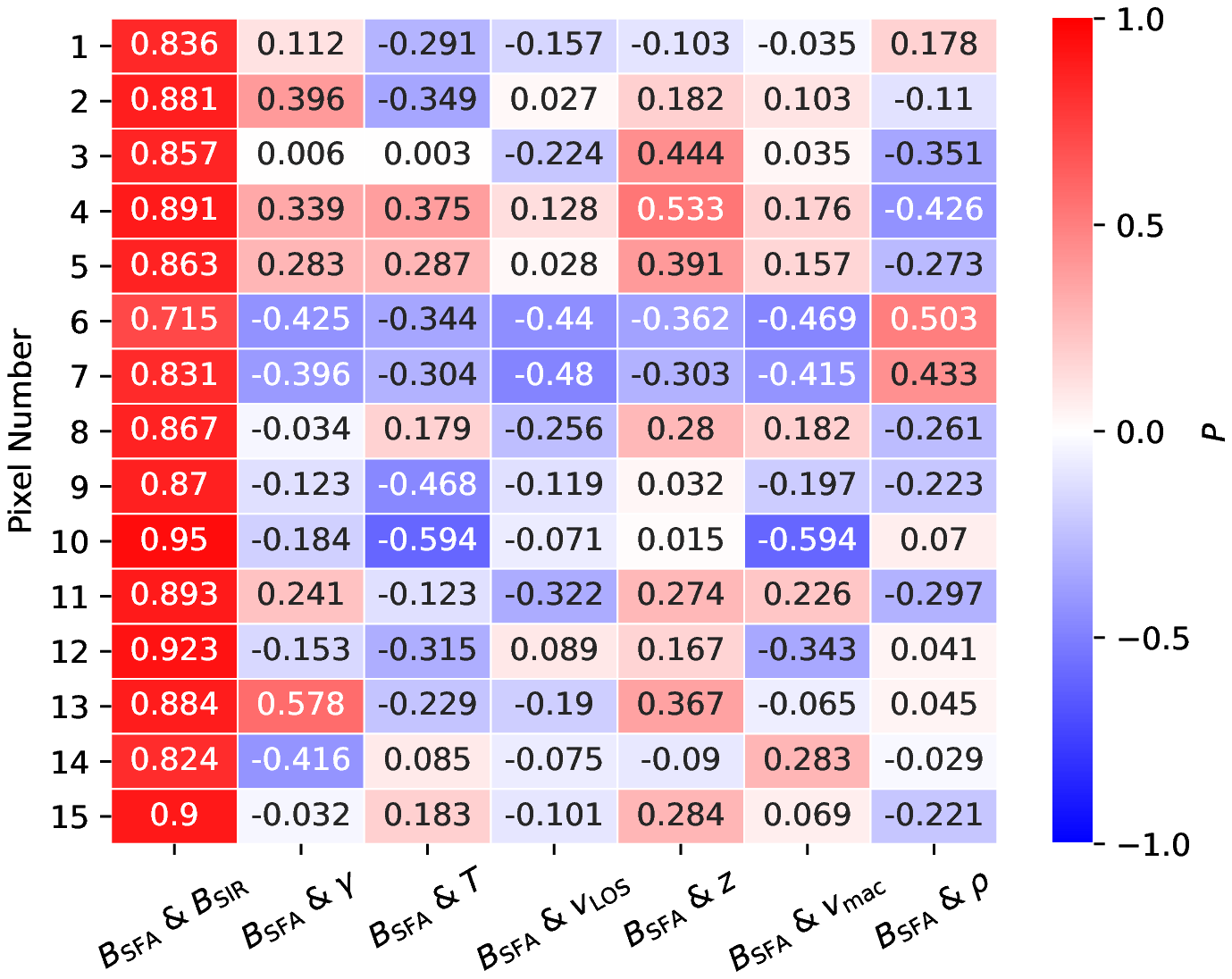}
    
    \includegraphics[width=\columnwidth]{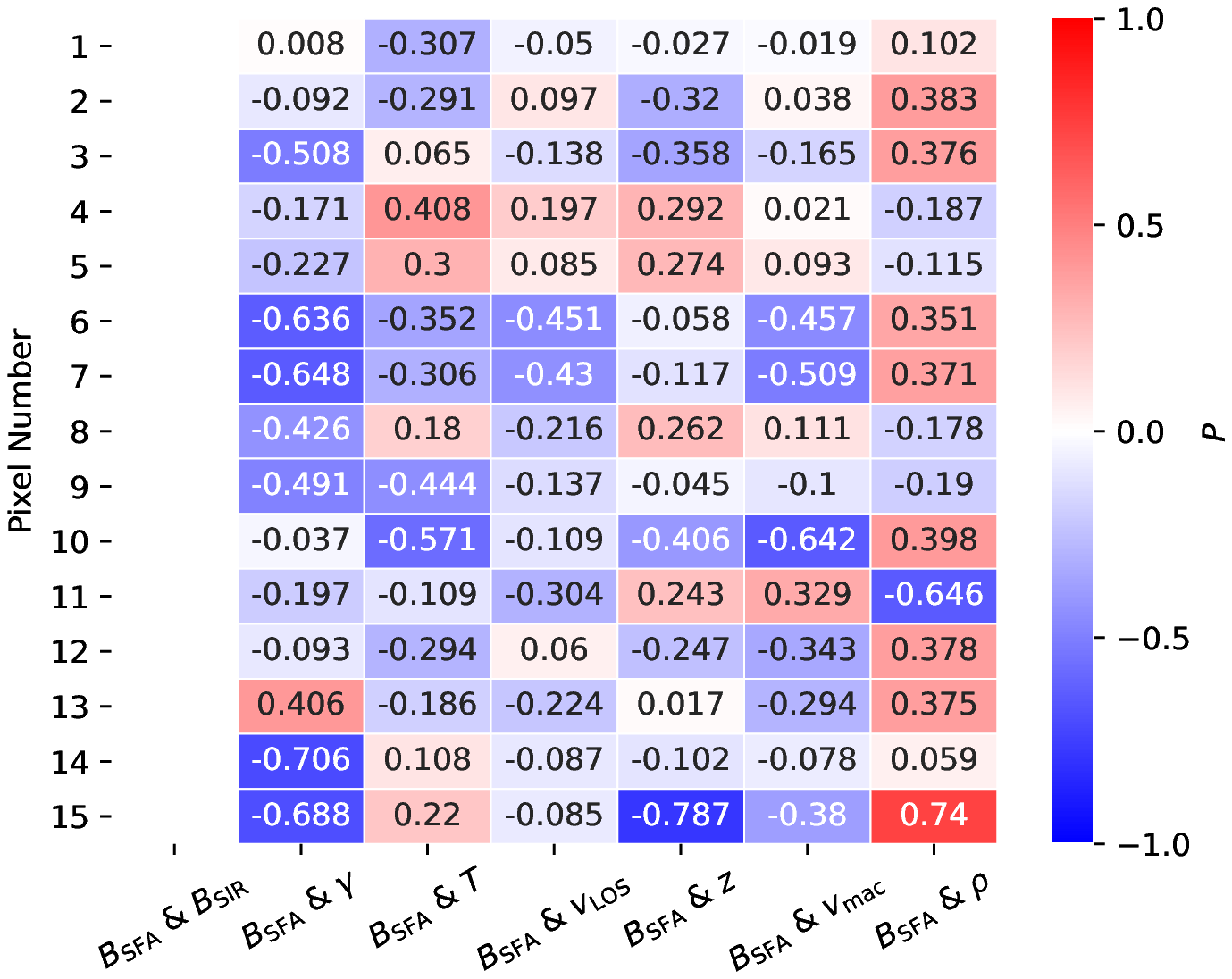}
    \caption{(Top panel) Pearson correlation coefficients, $P$, between $B_\mathrm{SFA}$ and the values returned by the S1 inversions for every pixel in Table \ref{Stats_GRIS}. A value of $1$ indicates perfect positive correlation $(red)$, while a value of $-1$ indicates perfect negative correlation $(blue)$. A value of $0$ represents complete uncorrelation. (Bottom panel) Same but for S2 inversions, where $B_\mathrm{SIR}$ was not included as a free parameter. The correlation between $B_\mathrm{SFA}$ and $B_\mathrm{SIR}$ is, therefore, undefined.}
    \label{fig:Pearson_original}
\end{figure}

Rerunning the inversions using S2 allows us to explore whether other processes (e.g. changes in the optical depth or bending of the magnetic field lines) could account for these changes by removing the dominant influence of the LOS magnetic field strength on the SIR inversions. For S2, $B$ was calculated at the first time-step in the same manner as S1, before the number of nodes in $B$ was set to zero in the subsequent frames, so that this parameter could not change. The blue line in the top left panel clearly displays that $B_\mathrm{SIR}$ does not change through time in the S2 inversions for Pixel 15. The values of $T$, $v_{\mathrm{LOS}}$, and $v_{\mathrm{mac}}$ are essentially invariant between S1 and S2 indicating that these parameters are likely not responsible for the observed changes in the Stokes vector through time. Additionally, the $\chi^2$ value for S2 matches well with that returned during the S1 inversions implying that both schemes fit these data equally well. We do, however, observe apparently periodic variations in $z$ (of about $25$ km), $\gamma$ (of up to $8^{\circ}$), and $\rho$ (up to $0.45$ g km$^{-1}$) differentiating S2 from S1 for this pixel. Importantly, these variations appear to be out-of-phase (in-phase for $\rho$) with $B_\mathrm{SFA}$. We note that large differences between $z$ and $\gamma$ in S1 and S2 inversions are not observed for all pixels, with these parameters generally displaying similar behaviours through time in both schemes. 

The top panel of Fig. \ref{fig:Pearson_original} shows the Pearson correlation coefficients, $P$, calculated between $B_\mathrm{SFA}$ and the atmospheric parameters returned by the S1 inversion for all of the pixels reported in Table~\ref{Stats_GRIS}. Again, all atmospheric parameters are recorded at $log\tau_{5000}$=$-0.5$, except $T$ which is recorded at $log\tau_{5000}$=$0.5$. The very strong correlation between $B_\mathrm{SFA}$ and $B_\mathrm{SIR}$ is immediately evident for every pixel (always above $0.71$); however, beyond that, no strong correlation is detected between any of the variables returned by the inversions. The bottom panel of Fig. \ref{fig:Pearson_original} details the $P$ values calculated between $B_\mathrm{SFA}$ and the atmospheric parameters returned by S2. Of course, the $P$ value between $B_\mathrm{SFA}$ and $B_\mathrm{SIR}$ is not defined as $B_\mathrm{SIR}$ is invariant in this inversion scheme and, hence, that column remains unfilled. Generally, no strong correlation is found between $B_\mathrm{SFA}$ and any SIR parameter, although, some notable $P$ values are returned. For Pixel 15, we observe a relatively strong $P$ value of $-0.787$ between $B_\mathrm{SFA}$ and $z$, indicating that observing at different heights in the solar atmosphere through time may be responsible for the observed changes in $B_\mathrm{SFA}$. We also observe a potential correlation between $B_\mathrm{SFA}$ and $\rho$ for Pixel 15, as would be expected if we were observing at different heights in the atmosphere at different times. No other strong correlations are found between $B_\mathrm{SFA}$ and either $z$ or $\rho$ for any other pixels. A potential trend does exist with four pixels indicating some anti-correlation between $B_\mathrm{SFA}$ and $\gamma$ with $P$ values of less than $-0.6$. It must be stressed, though, that a $P$ value of $-0.6$ only indicates potential anti-correlation and, as such, no further conclusions can be drawn from this. 

Finally, we applied wavelet analysis to the $z$, $\gamma$, and $v_\mathrm{LOS}$ signatures for each of these 15 pixels, finding 2 (4), 2 (4), and 0 (0) pixels which display significance in the temporal range studied here for S1 (S2), respectively. For S1, only Pixel $13$ returned significance above the $99$ \% level for both $z$ and $\gamma$, whilst for S2, only Pixel $3$ displayed significance above the $99$ \% level for both $z$ and $\gamma$. Pixel $15$ (plotted in Fig.~\ref{fig:temporal_plots}) showed evidence for strong oscillations in $z$ in the temporal range $600$-$1100$ s in S2 but did not return any significance for $\gamma$. From S1, we are able to conclude, therefore, that changes in $B_\mathrm{SIR}$ at a depth of around $log\tau_{5000}$=$-0.5$, potentially caused by compression in the solar atmosphere, could account for the observed changes in the Stokes vector. For S2, combining the Pearson correlation coefficients and the wavelet analysis does not highlight any single atmospheric parameter returned which appears to be universally responsible for the observed oscillation in $B_\mathrm{SFA}$. However, our results are not able to rule out that changes in the optical depth and magnetic field inclination could potentially be important in driving the observed changes in the Stokes vectors.

\section{Discussion}
\label{Discussion}

The presence of $3$ and $5$ minute oscillations within pores has been well established over the past decades (see, for example, \citealt{Fujimura09, Morton11, Moreels13, Grant15, Freij16}). However, it is now becoming clear that a range of oscillations are also present with long-periods, above $10$ minutes (previous examples include \citealt{Freij16, Riehokainen19, Stangalini21}). Here, we analysed the behaviour of the LOS magnetic field strength inferred from the Stokes-{\it{V}} parameter around the \ion{Fe}{I} $15648.52$ \AA\ line ($g_\mathrm{eff}$ of $3$) using the SFA (see Fig.~\ref{SFA_GRIS}). We identified $15$ pixels (detailed in Table~\ref{Stats_GRIS}) within three localised ($>1$\arcsec$^2$) regions (red boxes over-laid on the left and centre panels of Fig.~\ref{Wavelet_GRIS}) which displayed strong evidence for periods between $600$ and $1272$ s as inferred using the wavelet technique (\citealt{Torrence98}). Cross-correlation applied between individual frames indicated that the pointing remained stable during these observations indicating that mis-alignments with time were not responsible for the variations in the Stokes vector (see Fig.~\ref{Average_GRIS}). Additionally, these oscillations were present across multiple tiles in the observation and at different positions on the tiles (i.e., not only at locations where tiles are stitched together) indicating that neither instrumental nor data calibration effects account for these signals. Finally, no correlated periodicity was present in the Fried's parameter or intensity contrast confirming that seeing conditions were not the predominant driver of these changes.

The most striking aspect of these oscillations is their apparent amplitudes, with peak-to-peak variations of up to $420$ G ($20$ \% of the background field value) being measured. These values, an order of magnitude larger than amplitudes returned for periodicities of $3$-$6$ minutes (\citealt{Fujimura09}), are difficult to explain using current models.  Assuming an Alfv\'en speed equal to the sound speed in the photosphere ($\sim10$ km s$^{-1}$) and a period of 1000 s, it can be calculated that a natural MHD oscillation should have a wavelength of around $10$ Mm, deeper than some sunspot models but not ruled out by others (\citealt{Moradi10}). Interestingly, both the spatial extents and periods of these oscillations do match those predicted for similar magnetic field strengths by the slow-wave subphotospheric resonator model (\citealt{Zhugzhda18}), however, whether this is more than just mere coincidence will need to be investigated in more detail in the future. Analysis of other pixels indicated that longer-period (more than $2000$ s in some cases) amplitude variations were also present (see Fig.~\ref{Variations_GRIS}), although, a longer time-series would need to be studied in the future to confirm this using signal analysis methods. 

Having ruled out instrumental and seeing effects, we used SIR inversions (\citealt{SIR}) to investigate the initial validity of two hypotheses about how these oscillations were driven. Our first hypothesis was that these oscillations in the LOS magnetic field strength were true changes in the magnetic field strength at a single height in the solar atmosphere caused by horizontal motion driven compression (\citealt{Ulrich96}). To test this, we ran the inversions with two nodes in the LOS magnetic field strength with height (S1 from Table~\ref{table:nodes}). These inversions accounted for the changes in the Stokes vector by varying the LOS magnetic field strength which led to strong correlations between $B_\mathrm{SFA}$ and $B_\mathrm{SIR}$ (see the first column of the top panel of Fig.~\ref{fig:Pearson_original} where $B_\mathrm{SIR}$ is plotted at an optical depth of $log\tau_{5000}$=$-0.5$). No strong correlation was found between $B_\mathrm{SFA}$ and any other variable returned by this inversion scheme indicating that real changes in the strength of the magnetic field at a given height in the solar atmosphere could account for the observed variations. The reproduction of the $5$ minute oscillations in $v_\mathrm{LOS}$ (with amplitudes of around $0.4$ km s$^{-1}$) give us confidence that our inversions are capturing at least some real aspects of the solar atmosphere. Elucidating the physical compression mechanism which could account for such large oscillations in the magnetic field strength is beyond the scope of this article.

Secondly, we tested whether changes in other parameters, specifically changes in the optical depth or inclination angle (associated with Alfv\'en wave driven bending of magnetic field lines; \citealt{Ulrich96}), $\gamma$, could account for the observed changes in the Stokes vector.  To do this, we re-ran the inversions with a fixed stratification in $B_\mathrm{SIR}$ (calculated using two nodes at the first time-step). Importantly, no clear relationship was found between $B_\mathrm{SFA}$ and any variable returned by the inversions (see, bottom panel of Fig.~\ref{fig:Pearson_original}). Regarding opacity effects, if we assume a value of $4$ G km$^{-1}$ for the stratification of the magnetic field in the lower solar atmosphere within the pore (agreeing with values reported by \citealt{Sutterlin98}) then modifications of around $100$ km to the observed geometric height would be required to account for a $400$ G oscillation. We did not find evidence for such drastic changes in our inversions with variations of $10$s of km being the upper limit (see Pixel 15 plotted in Fig.~\ref{fig:temporal_plots}). Regarding the inclination angle, the pore observed here was at $\mu$=$0.87$, giving a viewing angle of approximately $30^{\circ}$ from the normal. Assuming a LOS magnetic field strength of $1950$ G (as in Fig.~\ref{fig:temporal_plots}), simple trigonometry can then be used to obtain a `true' vertical magnetic field strength of approximately $2250$ G. An oscillation of $4^{\circ}$ in the inclination angle (from $26^{\circ}$ to $34^{\circ}$) could then vary the LOS magnetic field strength by approximately $160$ G. The combined effects of changes in the optical depth and inclination angle could clearly, therefore, lead to variations in the magnetic field strength of the order hundreds of G. However, only Pixel $15$ returned a potential relationship between $B_\mathrm{SFA}$ and these quantities in S2. 

Although we are not able to identify any single parameter in S2 which can account for the oscillations observed here, the recent detection of Alfv\'en waves in a pore with longer periods than those reported here (\citealt{Stangalini21}) implies that bending of the field lines over time-scales of $600$-$1272$ s is feasible in a pore of this size making this an interesting topic for future study. Overall, our results do not rule out any of the potential scenarios (changes in optical depth, bending, or compression) which could explain the observed changes in the Stokes vector, however, more research will be required before stronger statements can be made in this direction. Wavelet analysis applied to the $z$, $\gamma$, and $v_\mathrm{LOS}$ signatures did not reveal the presence of strong oscillations in any of these values for the majority of pixels for either the S1 or S2 inversion schemes.

\section{Conclusions}
\label{Conclusions}

In this article, we reported on the detection of an intriguing new class of oscillations in the deep photosphere of an evolving solar pore, namely extremely localised ($<1$\arcsec$^2$), high-amplitude (up to $210$ G), long-period ($>600$ s) oscillations apparent in the LOS magnetic field strength inferred using the SFA. Our analysis ruled out instrumental and seeing effects as the cause of the changes in the Stokes vector; however, we were unable to confidently conclude what mechanism in the solar atmosphere is responsible for these oscillations. Numerical inversions revealed that both real oscillations in the magnetic field strength, associated with compression in the photosphere caused by horizontal motions, as well as several other effects including changes in opacity or bending, variations in the inclination of the magnetic field potentially induced by Alfv\'en waves, could potentially reproduce the variations in the Stokes vectors. Future work will, therefore, be required to better understand whether these oscillations are present in other pores (as well as sunspots), what their relationship is to the zoo of other oscillations in these structures, and whether they truly are compression oscillations of the magnetic field strength. Specifically, combining data from several spectral windows obtained using multiple instruments at the Daniel K. Inouye Solar Telescope (DKIST; \citealt{Rimmele20}) will help us to better understand these signatures in the coming years by allowing us to probe the solar atmosphere at a much larger range of optical depths simultaneously.

\begin{acknowledgements}
The authors are indebted to C. Dominguez-Tagle, J. Rendtel, and M. Collados for their help with data collection and reduction. C.J.N. and M.M. are thankful to the Science and Technology Facilities Council (STFC) for the support received to conduct this research through grant numbers ST/P000304/1 \&  ST/T00021X/1. This research data leading to the results obtained has been supported by SOLARNET project that has received funding from the European Union’s Horizon 2020 research and innovation programme under grant agreement no 824135. R.J.C. acknowledges support from the Northern Ireland Department for the Economy (DfE) for the award of a PhD studentship. Wavelet software was provided by C. Torrence and G. Compo, and is available at URL: \url{http://atoc.colorado.edu/research/wavelets/}. The 1.5-meter GREGOR solar telescope was built by a German consortium under the leadership of the Leibniz-Institute for Solar Physics (KIS) in Freiburg with the Leibniz Institute for Astrophysics Potsdam, the Institute for Astrophysics G\"ottingen, and the Max Planck Institute for Solar System Research in Göttingen as partners, and with contributions by the Instituto de Astrof\'isica de Canarias and the Astronomical Institute of the Academy of Sciences of the Czech Republic. The redesign of the GREGOR AO and instrument distribution optics was carried out by KIS whose technical staff is gratefully acknowledged. SDO/HMI data are courtesy of NASA/SDO and the HMI science team.
\end{acknowledgements}

\bibliographystyle{aa}
\bibliography{GRIS_sunspot}

\begin{thebibliography}{50}
\expandafter\ifx\csname natexlab\endcsname\relax\def\natexlab#1{#1}\fi

\bibitem[{{Bahng}(1958)}]{Bahng58}
{Bahng}, J.~D.~R. 1958, \apj, 128, 145

\bibitem[{{Bakunina} {et~al.}(2013){Bakunina}, {Abramov-maximov}, {Nakariakov},
  {Lesovoy}, {Soloviev}, {Tikhomirov}, {Melnikov}, {Shibasaki}, {Nagovitsyn},
  \& {Averina}}]{Bakunina13}
{Bakunina}, I.~A., {Abramov-maximov}, V.~E., {Nakariakov}, V.~M., {et~al.}
  2013, \pasj, 65, S13

\bibitem[{{Beckers} \& {Tallant}(1969)}]{Beckers69}
{Beckers}, J.~M. \& {Tallant}, P.~E. 1969, \solphys, 7, 351

\bibitem[{{Bharti} {et~al.}(2020){Bharti}, {Sobha}, {Quintero Noda}, {Joshi},
  \& {Pandya}}]{Bharti20}
{Bharti}, L., {Sobha}, B., {Quintero Noda}, C., {Joshi}, C., \& {Pandya}, U.
  2020, \mnras, 493, 3036

\bibitem[{{Borrero} {et~al.}(2003){Borrero}, {Bellot Rubio}, {Barklem}, \& {del
  Toro Iniesta}}]{Borrero03}
{Borrero}, J.~M., {Bellot Rubio}, L.~R., {Barklem}, P.~S., \& {del Toro
  Iniesta}, J.~C. 2003, \aap, 404, 749

\bibitem[{{Campbell} {et~al.}(2021{\natexlab{a}}){Campbell}, {Mathioudakis},
  {Collados}, {Keys}, {Asensio Ramos}, {Nelson}, {Kuridze}, \&
  {Reid}}]{Campbell2021}
{Campbell}, R.~J., {Mathioudakis}, M., {Collados}, M., {et~al.}
  2021{\natexlab{a}}, arXiv e-prints, arXiv:2102.00942

\bibitem[{{Campbell} {et~al.}(2021{\natexlab{b}}){Campbell}, {Shelyag},
  {Quintero Noda}, {Mathioudakis}, {Keys}, \& {Reid}}]{Campbell21b}
{Campbell}, R.~J., {Shelyag}, S., {Quintero Noda}, C., {et~al.}
  2021{\natexlab{b}}, arXiv e-prints, arXiv:2107.01519

\bibitem[{{Centeno} {et~al.}(2005){Centeno}, {Socas-Navarro}, {Collados}, \&
  {Trujillo Bueno}}]{Centeno05}
{Centeno}, R., {Socas-Navarro}, H., {Collados}, M., \& {Trujillo Bueno}, J.
  2005, \apj, 635, 670

\bibitem[{{Cho} {et~al.}(2015){Cho}, {Bong}, {Nakariakov}, {Lim}, {Park},
  {Chae}, {Yang}, {Park}, \& {Yurchyshyn}}]{Cho15}
{Cho}, K.~S., {Bong}, S.~C., {Nakariakov}, V.~M., {et~al.} 2015, \apj, 802, 45

\bibitem[{{Collados} {et~al.}(2012){Collados}, {L{\'o}pez}, {P{\'a}ez},
  {Hern{\'a}ndez}, {Reyes}, {Calcines}, {Ballesteros}, {D{\'\i}az}, {Denker},
  {Lagg}, {Schlichenmaier}, {Schmidt}, {Solanki}, {Strassmeier}, {von der
  L{\"u}he}, \& {Volkmer}}]{Collados12}
{Collados}, M., {L{\'o}pez}, R., {P{\'a}ez}, E., {et~al.} 2012, Astronomische
  Nachrichten, 333, 872

\bibitem[{{Dorotovi{\v{c}}} {et~al.}(2014){Dorotovi{\v{c}}}, {Erd{\'e}lyi},
  {Freij}, {Karlovsk{\'y}}, \& {M{\'a}rquez}}]{Dorotovic14}
{Dorotovi{\v{c}}}, I., {Erd{\'e}lyi}, R., {Freij}, N., {Karlovsk{\'y}}, V., \&
  {M{\'a}rquez}, I. 2014, \aap, 563, A12

\bibitem[{{Felipe} \& {Khomenko}(2017)}]{Felipe17}
{Felipe}, T. \& {Khomenko}, E. 2017, \aap, 599, L2

\bibitem[{{Freij} {et~al.}(2016){Freij}, {Dorotovi{\v{c}}}, {Morton},
  {Ruderman}, {Karlovsk{\'y}}, \& {Erd{\'e}lyi}}]{Freij16}
{Freij}, N., {Dorotovi{\v{c}}}, I., {Morton}, R.~J., {et~al.} 2016, \apj, 817,
  44

\bibitem[{{Freij} {et~al.}(2014){Freij}, {Scullion}, {Nelson}, {Mumford},
  {Wedemeyer}, \& {Erd{\'e}lyi}}]{Freij14}
{Freij}, N., {Scullion}, E.~M., {Nelson}, C.~J., {et~al.} 2014, \apj, 791, 61

\bibitem[{{Fujimura} \& {Tsuneta}(2009)}]{Fujimura09}
{Fujimura}, D. \& {Tsuneta}, S. 2009, \apj, 702, 1443

\bibitem[{{Grant} {et~al.}(2015){Grant}, {Jess}, {Moreels}, {Morton},
  {Christian}, {Giagkiozis}, {Verth}, {Fedun}, {Keys}, {Van Doorsselaere}, \&
  {Erd{\'e}lyi}}]{Grant15}
{Grant}, S.~D.~T., {Jess}, D.~B., {Moreels}, M.~G., {et~al.} 2015, \apj, 806,
  132

\bibitem[{{Gri{\~n}{\'o}n-Mar{\'\i}n}
  {et~al.}(2020){Gri{\~n}{\'o}n-Mar{\'\i}n}, {Pastor Yabar}, {Socas-Navarro},
  \& {Centeno}}]{Grinon20}
{Gri{\~n}{\'o}n-Mar{\'\i}n}, A.~B., {Pastor Yabar}, A., {Socas-Navarro}, H., \&
  {Centeno}, R. 2020, \aap, 635, A64

\bibitem[{{Henriques} {et~al.}(2020){Henriques}, {Nelson}, {Rouppe van der
  Voort}, \& {Mathioudakis}}]{Henriques20}
{Henriques}, V. M.~J., {Nelson}, C.~J., {Rouppe van der Voort}, L. H.~M., \&
  {Mathioudakis}, M. 2020, \aap, 642, A215

\bibitem[{{Hirzberger} {et~al.}(2002){Hirzberger}, {Bonet}, {Sobotka},
  {V{\'a}zquez}, \& {Hanslmeier}}]{Hirzberger02}
{Hirzberger}, J., {Bonet}, J.~A., {Sobotka}, M., {V{\'a}zquez}, M., \&
  {Hanslmeier}, A. 2002, \aap, 383, 275

\bibitem[{{Hollweg}(1979)}]{Hollweg79}
{Hollweg}, J.~V. 1979, \solphys, 62, 227

\bibitem[{{Howard} {et~al.}(1968){Howard}, {Tanenbaum}, \& {Wilcox}}]{Howard68}
{Howard}, R., {Tanenbaum}, A.~S., \& {Wilcox}, J.~M. 1968, \solphys, 4, 286

\bibitem[{{Jess} {et~al.}(2020){Jess}, {Snow}, {Houston}, {Botha}, {Fleck},
  {Krishna Prasad}, {Asensio Ramos}, {Morton}, {Keys}, {Jafarzadeh},
  {Stangalini}, {Grant}, \& {Christian}}]{Jess20}
{Jess}, D.~B., {Snow}, B., {Houston}, S.~J., {et~al.} 2020, Nature Astronomy,
  4, 220

\bibitem[{{Keys} {et~al.}(2018){Keys}, {Morton}, {Jess}, {Verth}, {Grant},
  {Mathioudakis}, {Mackay}, {Doyle}, {Christian}, {Keenan}, \&
  {Erd{\'e}lyi}}]{Keys18}
{Keys}, P.~H., {Morton}, R.~J., {Jess}, D.~B., {et~al.} 2018, \apj, 857, 28

\bibitem[{{Khomenko} \& {Collados}(2015)}]{Khomenko15}
{Khomenko}, E. \& {Collados}, M. 2015, Living Reviews in Solar Physics, 12, 6

\bibitem[{{Kleint} {et~al.}(2020){Kleint}, {Berkefeld}, {Esteves}, {Sonner},
  {Volkmer}, {Gerber}, {Kr{\"a}mer}, {Grassin}, \& {Berdyugina}}]{Kleint20}
{Kleint}, L., {Berkefeld}, T., {Esteves}, M., {et~al.} 2020, \aap, 641, A27

\bibitem[{{Kolotkov} {et~al.}(2017){Kolotkov}, {Smirnova}, {Strekalova},
  {Riehokainen}, \& {Nakariakov}}]{Kolotkov17}
{Kolotkov}, D.~Y., {Smirnova}, V.~V., {Strekalova}, P.~V., {Riehokainen}, A.,
  \& {Nakariakov}, V.~M. 2017, \aap, 598, L2

\bibitem[{{Krishna Prasad} {et~al.}(2016){Krishna Prasad}, {Jess}, {Jain}, \&
  {Keys}}]{Krishna16}
{Krishna Prasad}, S., {Jess}, D.~B., {Jain}, R., \& {Keys}, P.~H. 2016, \apj,
  823, 45

\bibitem[{{Lemen} {et~al.}(2012){Lemen}, {Title}, {Akin}, {Boerner}, {Chou},
  {Drake}, {Duncan}, {Edwards}, {Friedlaender}, {Heyman}, {Hurlburt}, {Katz},
  {Kushner}, {Levay}, {Lindgren}, {Mathur}, {McFeaters}, {Mitchell}, {Rehse},
  {Schrijver}, {Springer}, {Stern}, {Tarbell}, {Wuelser}, {Wolfson}, {Yanari},
  {Bookbinder}, {Cheimets}, {Caldwell}, {Deluca}, {Gates}, {Golub}, {Park},
  {Podgorski}, {Bush}, {Scherrer}, {Gummin}, {Smith}, {Auker}, {Jerram},
  {Pool}, {Soufli}, {Windt}, {Beardsley}, {Clapp}, {Lang}, \&
  {Waltham}}]{Lemen12}
{Lemen}, J.~R., {Title}, A.~M., {Akin}, D.~J., {et~al.} 2012, \solphys, 275, 17

\bibitem[{{Lites} {et~al.}(1982){Lites}, {White}, \& {Packman}}]{Lites82}
{Lites}, B.~W., {White}, O.~R., \& {Packman}, D. 1982, \apj, 253, 386

\bibitem[{{Moradi} {et~al.}(2010){Moradi}, {Baldner}, {Birch}, {Braun},
  {Cameron}, {Duvall}, {Gizon}, {Haber}, {Hanasoge}, {Hindman}, {Jackiewicz},
  {Khomenko}, {Komm}, {Rajaguru}, {Rempel}, {Roth}, {Schlichenmaier},
  {Schunker}, {Spruit}, {Strassmeier}, {Thompson}, \& {Zharkov}}]{Moradi10}
{Moradi}, H., {Baldner}, C., {Birch}, A.~C., {et~al.} 2010, \solphys, 267, 1

\bibitem[{{Moreels} {et~al.}(2015){Moreels}, {Freij}, {Erd{\'e}lyi}, {Van
  Doorsselaere}, \& {Verth}}]{Moreels15}
{Moreels}, M.~G., {Freij}, N., {Erd{\'e}lyi}, R., {Van Doorsselaere}, T., \&
  {Verth}, G. 2015, \aap, 579, A73

\bibitem[{{Moreels} {et~al.}(2013){Moreels}, {Goossens}, \& {Van
  Doorsselaere}}]{Moreels13}
{Moreels}, M.~G., {Goossens}, M., \& {Van Doorsselaere}, T. 2013, \aap, 555,
  A75

\bibitem[{{Morton} {et~al.}(2011){Morton}, {Erd{\'e}lyi}, {Jess}, \&
  {Mathioudakis}}]{Morton11}
{Morton}, R.~J., {Erd{\'e}lyi}, R., {Jess}, D.~B., \& {Mathioudakis}, M. 2011,
  \apjl, 729, L18

\bibitem[{{Riehokainen} {et~al.}(2019){Riehokainen}, {Strekalova}, {Solov'ev},
  {Smirnova}, {Zhivanovich}, {Moskaleva}, \& {Varun}}]{Riehokainen19}
{Riehokainen}, A., {Strekalova}, P., {Solov'ev}, A., {et~al.} 2019, \aap, 627,
  A10

\bibitem[{{Rimmele} {et~al.}(2020){Rimmele}, {Warner}, {Keil}, {Goode},
  {Kn{\"o}lker}, {Kuhn}, {Rosner}, {McMullin}, {Casini}, {Lin}, {W{\"o}ger},
  {von der L{\"u}he}, {Tritschler}, {Davey}, {de Wijn}, {Elmore}, {Fehlmann},
  {Harrington}, {Jaeggli}, {Rast}, {Schad}, {Schmidt}, {Mathioudakis},
  {Mickey}, {Anan}, {Beck}, {Marshall}, {Jeffers}, {Oschmann}, {Beard},
  {Berst}, {Cowan}, {Craig}, {Cross}, {Cummings}, {Donnelly}, {de Vanssay},
  {Eigenbrot}, {Ferayorni}, {Foster}, {Galapon}, {Gedrites}, {Gonzales},
  {Goodrich}, {Gregory}, {Guzman}, {Guzzo}, {Hegwer}, {Hubbard}, {Hubbard},
  {Johansson}, {Johnson}, {Liang}, {Liang}, {McQuillen}, {Mayer}, {Newman},
  {Onodera}, {Phelps}, {Puentes}, {Richards}, {Rimmele}, {Sekulic}, {Shimko},
  {Simison}, {Smith}, {Starman}, {Sueoka}, {Summers}, {Szabo}, {Szabo},
  {Wampler}, {Williams}, \& {White}}]{Rimmele20}
{Rimmele}, T.~R., {Warner}, M., {Keil}, S.~L., {et~al.} 2020, \solphys, 295,
  172

\bibitem[{{Ruiz Cobo} \& {del Toro Iniesta}(1992)}]{SIR}
{Ruiz Cobo}, B. \& {del Toro Iniesta}, J.~C. 1992, \apj, 398, 375

\bibitem[{{Scherrer} {et~al.}(2012){Scherrer}, {Schou}, {Bush}, {Kosovichev},
  {Bogart}, {Hoeksema}, {Liu}, {Duvall}, {Zhao}, {Title}, {Schrijver},
  {Tarbell}, \& {Tomczyk}}]{Scherrer12}
{Scherrer}, P.~H., {Schou}, J., {Bush}, R.~I., {et~al.} 2012, \solphys, 275,
  207

\bibitem[{{Schmidt} {et~al.}(2012){Schmidt}, {von der L{\"u}he}, {Volkmer},
  {Denker}, {Solanki}, {Balthasar}, {Bello Gonzalez}, {Berkefeld}, {Collados},
  {Fischer}, {Halbgewachs}, {Heidecke}, {Hofmann}, {Kneer}, {Lagg}, {Nicklas},
  {Popow}, {Puschmann}, {Schmidt}, {Sigwarth}, {Sobotka}, {Soltau}, {Staude},
  {Strassmeier}, \& {Waldmann }}]{Schmidt12}
{Schmidt}, W., {von der L{\"u}he}, O., {Volkmer}, R., {et~al.} 2012,
  Astronomische Nachrichten, 333, 796

\bibitem[{{Simon} \& {Weiss}(1970)}]{Simon70}
{Simon}, G.~W. \& {Weiss}, N.~O. 1970, \solphys, 13, 85

\bibitem[{{Sobotka}(2003)}]{Sobotka03}
{Sobotka}, M. 2003, Astronomische Nachrichten, 324, 369

\bibitem[{{Sobotka} {et~al.}(2013){Sobotka}, {{\v{S}}vanda},
  {Jur{\v{c}}{\'a}k}, {Heinzel}, {Del Moro}, \& {Berrilli}}]{Sobotka13}
{Sobotka}, M., {{\v{S}}vanda}, M., {Jur{\v{c}}{\'a}k}, J., {et~al.} 2013, \aap,
  560, A84

\bibitem[{{Solanki}(2003)}]{Solanki03}
{Solanki}, S.~K. 2003, \aapr, 11, 153

\bibitem[{{Stangalini} {et~al.}(2021){Stangalini}, {Erd{\'e}lyi}, {Boocock},
  {Tsiklauri}, {Nelson}, {Del Moro}, {Berrilli}, \&
  {Kors{\'o}s}}]{Stangalini21}
{Stangalini}, M., {Erd{\'e}lyi}, R., {Boocock}, C., {et~al.} 2021, Nature
  Astronomy

\bibitem[{{Stangalini} {et~al.}(2012){Stangalini}, {Giannattasio}, {Del Moro},
  \& {Berrilli}}]{Stangalini12}
{Stangalini}, M., {Giannattasio}, F., {Del Moro}, D., \& {Berrilli}, F. 2012,
  \aap, 539, L4

\bibitem[{{Suetterlin}(1998)}]{Sutterlin98}
{Suetterlin}, P. 1998, \aap, 333, 305

\bibitem[{{Torrence} \& {Compo}(1998)}]{Torrence98}
{Torrence}, C. \& {Compo}, G.~P. 1998, Bulletin of the American Meteorological
  Society, 79, 61

\bibitem[{{Ulrich}(1996)}]{Ulrich96}
{Ulrich}, R.~K. 1996, \apj, 465, 436

\bibitem[{{Wittmann}(1969)}]{Wittmann69}
{Wittmann}, A. 1969, \solphys, 7, 366

\bibitem[{{Zhugzhda} \& {Sych}(2018)}]{Zhugzhda18}
{Zhugzhda}, Y. \& {Sych}, R. 2018, Research in Astronomy and Astrophysics, 18,
  105

\bibitem[{{Zhugzhda} \& {Sych}(2014)}]{Zhugzhda14}
{Zhugzhda}, Y.~D. \& {Sych}, R.~A. 2014, Astronomy Letters, 40, 576

\end{thebibliography}

\end{document}